\documentclass[12pt]{article}

\advance\hoffset by -1.1truecm
\advance\voffset by -1.0truecm

\def\be{\begin{equation}}
\def\ee{\end{equation}}
\def\ba{\begin{eqnarray}}
\def\ea{\end{eqnarray}}

\newcommand{\R}{\mbox{I \hspace{-0.82em} R}}
\newcommand{\x}{{\bf x}}
\newcommand{\p}{{\bf p}}
\newcommand{\sn}{\smallskip\newline}
\newcommand{\mn}{\medskip\newline}

\begin{document}
\title{On Symmetric Operators in Noncommutative 
Geometry\thanks{Invited Contribution to the `Extended
Proceedings of the I.S.I. Guccia Workshop
on Quantum Groups, Noncommutative Geometry 
and Fundamental Interactions',
Eds. D. Kastler, M. Rosso, to appear in Nova Science 
Publ. (1999)}}
\author{Achim Kempf\\
Institute for Fundamental Theory, Department of Physics\\
University of Florida, Gainesville, FL 32611, USA} 

\date{}
\maketitle

\vskip-7.5truecm

\hskip11.7truecm
{\tt IFT-HEP-98-35} 

\hskip11.7truecm
{\tt math-ph/9811027}
\vskip6.5truecm

\begin{abstract}
In Noncommutative Geometry, as in quantum theory, 
classically real variables are 
assumed to correspond to self-adjoint operators. We 
consider the 
relaxation of the requirement  of self-adjointness 
to mere symmetry for operators $X_i$
which encode space-time information. 
\end{abstract}

\section{Introduction}
Part of the `dictionary' of Noncommutative
 Geometry \cite{connes} is the assumption that 
classically real variables correspond to 
self-adjoint operators in the full theory,
 in line with
the usual quantum mechanical description of 
observables through self-adjoint operators.
The use of self-adjoint operators is, of course, in many 
ways natural since, for example, they often generate 
unitary groups of physical significance.

On the other hand, there are indications that
 the structure of
space-time at distances close to the Planck scale
is highly nontrivial, see e.g. \cite{ps},
 and \cite{nc}.
In particular, as we will describe below 
in more detail, there
 is evidence which points towards
  a short-distance structure of space-time
which is beyond what can be described by 
self-adjoint operators, namely beyond continua and
lattices. 

Let us, therefore, reconsider the basic 
assumption that operators $X_i$ which encode space-time information 
within a fundamental theory of quantum gravity are self-adjoint.

Of course, if the fundamental theory of quantum gravity 
is linear as a quantum theory, i.e. if it obeys a linear
 superposition principle, then it is
natural to assume that it encodes space-time 
information through operators 
$X_i$ which are  \it linear. \rm And further, it 
is natural
to assume that all formal expectation values 
of these operators $X_i$ are real - in order 
to allow
some form of physical interpretation in terms 
of space-time coordinates. However, these
assumptions imply only that these operators
 $X_i$ are symmetric on their 
domain $D$. Self-adjointness, as is well known, 
is a stronger condition than symmetry.
We are therefore led to consider the possibility 
that the fundamental theory of quantum gravity
may encode space-time information using operators
 $X_i$ which are symmetric and not self-adjoint.
\sn
Clearly, if we wish to consider merely symmetric rather than
self-adjoint operators
 we will have to address two crucial questions:
\begin{enumerate}
\item  Which are the implications of the fact that 
generic symmetric operators, a priori, do not
 generate unitary groups?
\item Physically, which types of spatial structure 
are described by merely symmetric linear 
operators $X_i$ ?
\end{enumerate}

\section{Preliminaries}
Let us begin by briefly recalling von Neumann's 
theory of self-adjoint extensions
of symmetric operators.

Consider the M{\"o}bius transform $x \rightarrow
 (x-i)(x+i)^{-1}$, which maps the real line
onto the unit circle. An analogous transformation,
 the so-called Cayley transform,
\be
X \rightarrow S := (X-i1)(X+i1)^{-1}
\ee
maps the set of self-adjoint operators $X$ onto 
the set of unitaries. The Cayley transform does
this bijectively. Because of the bijectivity  
the Cayley transform is technically
 to be preferred over the exponentiation as a method to
 relate self-adjoints to unitaries.

For clarity of the argument, let us now consider 
the clear-cut case of an operator $X$ which is 
simple symmetric, i.e. which is symmetric but not 
self-adjoint and 
which is also not self-adjoint on any invariant sub space.
Recall that without restricting 
generality we can assume $X$ to be closed.

The Cayley transform maps $X$ into an 
operator $S$ which is merely isometric. This is
because either the domain or the range, or 
both, of $S$ are not the full Hilbert space 
$H$ - otherwise
$X$ would be self-adjoint.

The orthogonal complements of the domain 
and the range 
\be
L_+ := ((X+i1).D)^\perp
\ee
and
\be
L_- := ((X-i1).D)^\perp
\ee
are the so-called deficiency spaces. Their dimensions
 $(r_+,r_-)$ are the so-called
deficiency indices. Let us assume them to be finite. 
If both are equal, $r_+=r_-$, then it is possible to
 supplement $S$ by a finite rank isometry 
$S^\prime : ~ L_+ \rightarrow L_-$. In this way, $S$ 
is extended by $S^\prime$ to a unitary
operator $U:~ H\rightarrow H$. The inverse Cayley 
transform of $U$ yields a self-adjoint extension
$X_e$ of the simple symmetric $X$. Of course, the 
choice of $S^\prime$ is arbitrary and nonunique
up to elements of the group $U(r)$ which maps, say,
 $L_+$ into itself, where $r=r_+=r_-$ 
is the deficiency index.
Therefore, the set of self-adjoint extensions forms 
itself a representation of a $U(r)$. We will find
this group reappearing in our discussion later.

It is clear that a simple symmetric operator 
with unequal deficiency indices cannot have
self-adjoint extensions, since its isometric Cayley 
transform does not have unitary extensions.

\section{Symmetric Operators and Unitary Groups}

Let us now prepare the ground for answering the 
first of the questions which
we raised in the introduction:

A merely symmetric operator cannot be exponentiated 
to yield a unitary. Its exponentiation only yields
isometries, and similarly, as we recalled in the 
previous section, also the Cayley transform
yields mere isometries. 

Does this mean that simple symmetric
 operators cannot 
play a significant role? Perhaps unexpectedly, 
the answer is that simple symmetric 
operators, at least those with
equal deficiency indices, are nevertheless very much
at the heart of the the algebra $B(H)$!

Let us first consider the simple but instructive 
example of the differential operator
$X= i\partial_\lambda$:
\sn
We consider this operator $X$  
on a dense domain $D\subset L^2([0,1])$, where all
 functions $\phi(\lambda) \in 
D$ are infinitely differentiable and obey the 
boundary condition $\phi(0)=\phi(1) = 0$. On $D$, 
the operator $X$ is simple
symmetric with deficiency indices $(1,1)$. 

The exponentiation $S(a) := \exp(-i a X)$, being
 the exponentiation of a differentiation operator, 
$S(a) = \exp(a \partial_\lambda)$, 
is of course a translation operator $S(a) \phi(\lambda) = \phi(\lambda +a)$.
However,  since $S(a)$  is translating
functions confined to an interval, $S(a)$ is not 
unitary, but is instead a mere isometry: 
$S(a)$ can only act
on functions who's support is such that $S(a)$ is
 not translating their support
 beyond the ends of the interval.

We can, however, extend the operator $X$ to a 
self-adjoint operator $X_u$, by extending its
domain to include 
functions with the boundary condition $\phi(0) 
= u ~\phi(1)$. Here, the choice of 
the phase $u\in U(1)$ labels the self-adjoint 
extension.
In this way, one arrives at operators 
$S_u(a):=\exp(-i a X_u)$ which are self-adjoint.
These translation operators do in fact translate 
arbitrary functions. Whatever part of the function
 is translated beyond the interval boundary, reappears
 from the other boundary into the interval - 
phase rotated by the phase $u$.

Let us now consider the set of all 
unitaries which can be generated by the self-adjoint
 extensions $X_u$.
Clearly, we can compose unitaries $S_u(a), 
S_{u^\prime}(a^\prime)$ from
 different self-adjoint extensions, i.e.
  different
boundary conditions, to obtain more unitaries. 
For example, we can translate
in one direction with one phase relation followed 
by a translation back by the same amount -
crucially, however, 
the forward and the backward translations need
 not have the same boundary conditions, i.e. they
need not be using the same self-adjoint extension.
For example, 
\be
T:= S_{u^\prime}(-a) S_u(a) 
\ee
for some arbitrary fixed $0 <a<1$ is a unitary 
which does not translate at all. Instead, it phase
rotates functions $\phi(\lambda)$ by the phase 
$(u^\prime)^{-1}u\in U(1)$
on the interval $(1-a,1)$ and acts
as the identity for that part of functions which
 is defined on the interval $(0,1-a)$.
It becomes clear that 
the self-adjoint extensions of $X$ can be used 
not only to translate functions.
In addition, by suitably combining unitaries 
$S_u(a)$ for different $u\in U(1)$ and 
$a\in (-1,1)$, we can arrive at unitaries which 
describe arbitrary local phase rotations
in $\lambda$-space!
\sn
The same argument goes through if we define a 
differential operator $X$ which acts on
$r$ copies of the interval. Then, its deficiency 
indices are $(r,r)$, and the only change to the
above discussion is that the boundary conditions 
of the self-adjoint extensions are now
determined by unitary matrices $u \in U(r)$,
 rather than simply by a phase:
\be
\phi_i(0) = \sum_{j=1}^r u_{ij} ~\phi_j(1)
\ee
We conclude that the exponentiated self-adjoint 
extensions of $X$ are able to generate 
arbitrary local $U(r)$- transformations - 
in addition to the translations which they
of course also generate. 
\mn
This means that the set of self-adjoint extensions 
of this simple symmetric $X$ with equal 
deficiency indices $r$ generate translations, 
accelerations, \it and \rm $r$-dimensional local 
iso-rotations. In fact, as will be shown in detail in
\cite{ak-sds}, the self-adjoint extensions always 
generate the whole of $B(H)$! In order to be
precise, let us now use the unitary extensions of
 the Cayley transform, as discussed in
section 2: The statement is that the weak closure
 of the
*-algebra generated by the
coset of all unitary extensions of the Cayley 
transform of a simple symmetric $X$ with 
finite equal deficiency indices, is $B(H)$.
Therefore, if simple symmetric $X$ emerge in a 
physical theory, they may be expected to
play a central r{\^o}le.

Indeed, as we will discuss at the end, there
automatically arises an $r$-dimensional
`isospinor' structure if the deficiency 
indices are $(r,r)$.

\section{Symmetry and self-adjointness}
Let us now prepare the ground for answering 
the second of the questions raised in the
introduction.

In order to clarify in which ways space-times 
described by simple symmetric operators would differ
from space-times described by self-adjoint 
operators, let us recall 
from section 2 that simple symmetric operators 
can be categorized by their
deficiency indices into two classes. 
The $X$ with finite, equal deficiency indices
 $(r,r)$
have a $U(r)$- family of self-adjoint extensions, 
and those with unequal finite 
deficiency indices  do not have self-adjoint 
extensions.

The spatial short-distance structures described 
by simple symmetric operators were 
named `fuzzy' in \cite{ak-bialo}. The particular
 subclasses of operators of equal and unequal
deficiency indices were denoted fuzzy-A 
and fuzzy-B, respectively, in \cite{ak-erice}.
There, it was also first described that 
there exists a physically more intuitive,
equivalent definition of the fuzzy-A case:
\sn
$X$ is simple symmetric with equal deficiency 
indices exactly iff there exists a 
positive function 
\be 
\Delta X_{\mbox{\tiny min}}(\xi) > 0,
\ee
so that for each $\xi \in \R$, 
all normalized $ \vert \phi \rangle \in D$ 
with expectation
$\langle \phi \vert X \vert \phi \rangle = \xi$ obey
\be
(\Delta X)_{\vert \phi \rangle} \ge \Delta
 X_{\mbox{\tiny min}}(\xi) .
\ee
Using this characterization, we can now describe operators of the type fuzzy-A 
as yielding spaces in which the standard 
deviation in positions - we are here formally using
quantum mechanical terminology - 
could not be made smaller than some 
finite lower bound $\Delta X_{\mbox{\tiny
 min}}(\xi)$. Since the lower bound is in general some
function of the expectation value $\xi$
the amount of `fuzzyness' can in general 
vary from place to place.
\sn
The fuzzy-B case, i.e. the case simple symmetric 
operators $X$ with unequal deficiency indices
can be shown to describe short distance structures
which are such that  there exist sequences of 
vectors in the physical domain such that $\Delta 
x$ converges to zero.
The fuzzy-B type short-distance structures are
 `fuzzy' in the sense 
that vectors of increasing localization 
around different expectation values then in 
general do not become orthogonal.
\sn
Among studies into the structure of space-time 
at the Planck scale there appeared
 evidence
for effective correction terms to the uncertainty 
relation, in the simplest case
\be
\Delta x \Delta p \ge \frac{\hbar}{2}\left( 1 + 
\beta ~(\Delta p)^2 + ... \right) \label{ucr}
\ee
for some $\beta>0$, related either to the Planck 
scale or to a string scale.
For a sufficiently small constant 
$\beta$, the correction term is negligible at
 present-day experimentally accessible
scales.  The
correction term nevertheless implies a crucial 
new feature, namely
that $\Delta x$ is now finitely bounded from below by  
\be
\Delta x_{\mbox{\tiny min}} ~  =  ~  \hbar~ \sqrt{\beta}
\ee
For reviews on the origins of this type of
 uncertainty relation, see e.g. \cite{garay,witten}.
\sn
It is clear from the physical characterization
 of the fuzzy-A case above, that 
any operator, within any theory, which obeys 
this type of uncertainty relation, is
simple symmetric with equal deficiency indices.

\section{Conclusions}
Let us now try to tie the discussed issues 
together, in order to answer the questions
raised in the beginning. We asked about the
 mathematical and physical implications of 
relaxing the `dictionary entry' for real variables 
from self-adjointness to mere symmetry.
We have discussed some aspects already, but 
it remains to address the question 
which r{\^o}le unitaries and isometries may
 play in a theory with simple symmetric $X_i$.
To this end we will also have to address the
 question how,  in practice,
a physical theory may determine that
its operators $X_i$ are merely symmetric
 rather than self-adjoint.
\sn
Let us begin by noting that a
generic symmetric operator may be self-adjoint
 and simple symmetric on different subspaces.
This means that it can describe arbitrary mixtures
 of the basic cases of short-distance structures,
namely continua, lattices and the two fuzzy cases.

We recall  that there are numerous physical 
systems in which the self-adjoint extensions
of various differential operators correspond
 physically to 
choices of boundary conditions - which are 
imposed externally onto the physical system.
The question arises, therefore,  whether a 
theory can \it intrinsically \rm specify
 that an operator $X_i$ is
 simple symmetric - even if 
self-adjoint extensions exist in the Hilbert space.
To see that theories can easily yield such 
intrinsic domain specifications, we note that any
dynamical or kinematical operator equation
 within a theory involves a domain condition,
  in particular
if unbounded operators are involved. If a theory 
requires equations among its operators, then
a physical domain can only be a domain on which
 all of the operators are well-defined.
\sn
For example, the stringy uncertainty relation 
above, can be induced by the commutation
relation $[\x,\p] = i\hbar (1 +\beta \p^2)$. It
 is clear that in any domain $D$ in a Hilbert space
$H$, on which this equation holds, the operator $X$
can only be simple symmetric, as is easily concluded from the implied
 uncertainty relation. This is in spite of $\x$ having
self-adjoint extensions in the Hilbert space. We remark that 
the functional analysis of generalized uncertainty
relations of this type was first considered 
in \cite{ak-ucr}.  
\sn
In this way, any dynamical or kinematical equation
 between the $X_i$ 
and arbitrary other operators within a fundamental theory of 
quantum gravity can induce domain specification for 
the operators $X_i$. Any such theory-intrinsic
domain specification
could determine that certain $X_i$ are simple symmetric and 
cannot be extended within
the physical domain, i.e. within the domain on which
 the theory's equation hold. Those
dynamical or kinematical equations could of course also specify 
unequal deficiency indices.  Domain specifications and possibly
a noncommutativity of the $X_i$ 
may also arise with path integrals. In this context, see \cite{gm1,gm2}. 

Of course, the structure of space-time may in general vary 
arbitrarily from place to place in particular, in an in general 
$n$ dimensional
noncommutative space. This is here reflected by the fact
 that  the functional
analysis of any one $X_i$ is a function of the 
analysis of the other $X_i$.
\mn
We can now finally address the question of 
isometries and unitaries:
\sn
Simple symmetric operators, though densely defined 
cannot be defined on the entire Hilbert space,
since they are discontinuous operators, i.e. they are
blind to part of the Hilbert space. 
As we have seen in the case of equal deficiency
 indices, all vectors which would 
describe structure smaller than a finite
smallest uncertainty are cutoff from the domain.
Unitaries and isometries, however, are bounded
 operators which by continuity can see
all of the Hilbert space. No part of the Hilbert 
space can be hidden from them. 
This indeed yields
a mechanism by which the cutoff degrees of freedom
 reappear, producing an isospinor
structure of internal degrees of freedom:
\sn
Let us consider the case of a single $X$ with equal
 deficiency indices in more detail:
For each real $\xi$ there can  be shown
 \cite{ak-sds} to exist a self-adjoint extension
in which $\xi$ is an $r$-fold degenerate 
eigenvalue. Thus, each real $\xi$
is an $r$-fold eigenvalue of the adjoint $X^*$ 
with eigenvector say $\vert \xi,i\rangle$,
where $i=1,2,...,r$. Using these, any vector can
 be expanded in an `isospinor function' 
$\phi_i(\xi) := \langle \xi,i\vert \phi\rangle$. 
At large distances the eigenvectors become orthogonal,
yielding an ordinary isospinor function. 
From a cutoff with deficiency indices $(r,r)$
an isospinor structure and a corresponding 
local $U(r)$
structure have emerged automatically!

A precise definition of `gauge transformations',
 can be given which indeed yields ordinary local gauge transformations
at large distances: the idea is to define
gauge transformations as 
isometries which commute with the $X$ and which 
map from a physical domain into a 
physical domain. 
The emergence of gauge transformations in this 
way was first discussed in
\cite{ak-euro}. A more detailed study, 
including also the case of unequal
deficiency indices, will given in \cite{ak-sds}.
\mn
To summarize, we conclude that the mathematics 
of symmetric versus 
self-adjoint and isometric versus unitary 
operators offers new possibilities
for the description of space-time short-distance
 structures within theories which
are linear as quantum theories. In particular, 
there emerges
a new mechanism by which ultraviolet cutoff
degrees of freedom can reappear by inducing an
 isospinor structure of internal degrees of
freedom. 

It should be most interesting to study 
potential applications of the tools of
 noncommutative geometry in this context,
in particular, as cohomological methods tend to
focus on larger scale structures.

Finally, we remark that in \cite{broutetal} the fuzzy-A type 
short-distance structure has recently been applied to 
the resolution of the transplanckian energy problem of 
black hole radiation.


\begin{thebibliography}{99}
\bibitem{connes} A. Connes, \it Noncommutative
 Geometry, \rm AP (1994)

\bibitem{ps} S.W. Hawking, Nucl.Phys. {\bf 
B144}, 349 (1978), ~~ 
 D.J. Gross, 
P.F. Mende, Nucl. Phys. {\bf B303}, 407 (1988),~~
D. Amati, M. Ciafaloni, G. Veneziano, Phys.Lett. {\bf B216} 
41, (1989),~~
S. Doplicher, K. Fredenhagen, J.E. Roberts,
Comm. Math. Phys. {\bf 172}, 187 (1995),~~
D.V. Ahluwalia, Phys. Lett. {\bf B339}, 301 (1994),~~
G. Amelino-Camelia, John Ellis, N.E.
 Mavromatos, D.V. Nanopoulos,
Mod.Phys.Lett.{\bf A12} 2029 (1997),~~
M.-J. Jaeckel, S. Reynaud, Phys. Lett. 
{\bf A185}, 143 (1994),~~
C. Rovelli, Preprint C-97-12-16, gr-qc/9803024, ~~
A. Jevicki, T. Yoneya, hep-th/9805069, ~~
S. de Haro, gr-qc/9806028

\bibitem{nc} A. Connes, M.R. Douglas, A. Schwarz, RU-97-94,
 hep-th/9711162, ~~ 
J. Madore, \it An introduction to noncommutative 
differential geometry and its physical applications,
 \rm  CUP (1995), ~~
S. Majid, \it Foundations of Quantum Group Theory, 
\rm CUP (1996),~~
A.H. Chamseddine, A. Connes Phys.Rev.Lett.{\bf 77},
 4868 (1996)

\bibitem{ak-sds} A. Kempf, in preparation

\bibitem{ak-bialo} A. Kempf, to appear in Rep. of 
Math. Phys. (1998),  hep-th/9806013

\bibitem{ak-erice} A. Kempf, to appear in Proc. 36th 
Erice School \it From the Planck Length
to the Hubble Radius, \rm Preprint UFIFT-HEP-98-30

\bibitem{garay} L.J. Garay, Int. J. Mod. 
Phys. {\bf A10}, 145 (1995)

\bibitem{witten} E. Witten, Phys. Today 
{\bf 49} (4), 24 (1996)

\bibitem{ak-ucr} A. Kempf, J. Math. Phys. 
{\bf 35} (9), 4483 (1994)

\bibitem{gm1} G. Mangano, J. Math. Phys. {\bf 39}, 2584 (1998), 
gr-qc/9705040 

\bibitem{gm2} G. Mangano, hep-th/9810174

\bibitem{ak-euro} A. Kempf, Europhys. Lett.
 {\bf 40} (3), 257 (1997),
hep-th/9706213 

\bibitem{broutetal} R. Brout, C. Gabriel, 
M. Lubo, P. Spindel, hep-th/9807063


\end{thebibliography}
\end{document}